\documentclass[conference]{IEEEtran}

\usepackage[T1]{fontenc}
\usepackage[utf8]{inputenc}
\usepackage{cite}
\usepackage{amsmath}
\usepackage{amssymb}
\usepackage{booktabs}
\usepackage{graphicx}
\usepackage{tikz}
\usepackage{pgfplots}
\usepackage{xspace}
\usepackage{url}
\usepackage{algorithm}
\usepackage{algpseudocode}

\usetikzlibrary{positioning,calc,patterns}
\pgfplotsset{compat=1.18}
\pgfplotsset{
ieeeplot/.style={
    width=\linewidth,
    axis line style={black, line width=0.7pt},
    tick style={black, line width=0.7pt},
    every axis label/.append style={font=\footnotesize},
    every tick label/.append style={font=\footnotesize},
    every axis plot/.append style={draw=black, line width=0.8pt},
    legend style={font=\footnotesize, draw=none, fill=none},
    nodes near coords style={font=\scriptsize},
    axis x line*=bottom,
    axis y line*=left
}}

\newcommand{\sys}{BoundSwitch\xspace}

\title{In-Network Artificial Computing Enhanced Light Model-Switching for Emergency Communications Networks}

\author{
\IEEEauthorblockN{
Yuehan Li\IEEEauthorrefmark{0}, 
Zhiyuan Ren\IEEEauthorrefmark{0}, 
Tao Zhang\IEEEauthorrefmark{0}, 
Wenchi Cheng\IEEEauthorrefmark{0}}
\IEEEauthorblockA{\IEEEauthorrefmark{0}School of Telecommunications Engineering, Xidian University, Xi'an 710071, China \\
E-mail: yhli225@stu.xidian.edu.cn, zyren@xidian.edu.cn, zhangtao02@xidian.edu.cn, wccheng@xidian.edu.cn
}
}
\begin{document}

\maketitle

\begin{abstract}
Emergency communications networks require in-network intelligence for timely traffic handling under dynamic demands and runtime constraints. In these environments, packets may need different inference behaviors, and conventional model replacement via control-plane updates is too slow for responsive operation. We propose an in-network artificial computing framework with lightweight model-switching, where multiple Binary Neural Network (BNN) models are kept resident within a shared execution framework. Packet metadata selects the active model at packet granularity with $O(1)$ selection cost. A fixed 1024-byte payload is aligned with x86 AVX-512, enabling efficient memory access. The framework is realized on an eBPF/XDP + AF\_XDP stack. Experimental results show that the system sustains 1.894~Mpps with a 0.528~$\mu$s inference latency, while model selection adds only 0.005~$\mu$s. Our results demonstrate that different resident models induce distinct packet-processing behaviors, that scaling to 16 slots preserves low switching overhead, and that online model switching completes without wrong-verdict packets. These results show the practicality of lightweight in-network artificial computing on commodity hardware.
\end{abstract}


\section{Introduction}

Emergency communications networks need both connectivity and timely in-network intelligent processing under dynamic demands and runtime constraints. In such environments, packets associated with different tasks may require different processing on the same forwarding path, making static in-network computing infeasible. Instead, inference behavior must adapt online in a lightweight and responsive manner.

This requirement exposes a systems challenge that is not well addressed by conventional model management approaches. In existing practice, a change in model behavior is often realized through control-plane-driven replacement, such as reloading weights, restarting services, or redirecting traffic to another execution instance. While such operations may be acceptable in relatively stable environments, they are too heavyweight for responsive online adaptation in emergency communications networks, where semantic requirements may shift at packet boundaries or over short service intervals. As a result, the key challenge is no longer only how to execute one model efficiently on the packet path, but how to support lightweight model-switching among multiple inference behaviors within a shared forwarding framework, without path mutation, service interruption, or online model delivery.

To address this challenge, we present an in-network artificial computing framework with enhanced light model-switching on the online packet path. The core idea is to keep multiple lightweight model behaviors continuously available within a shared execution framework and to resolve the active model through packet-carried metadata at packet granularity. In the current design, multiple Binary Neural Network (BNN) weight sets are preloaded into a resident model bank, while the parser, the executor, and the forwarding logic remain unchanged across packets. In this way, semantic adaptation is reduced to lightweight model resolution within one forwarding path, rather than being elevated into a heavyweight control-plane reconfiguration event. This design enables the same packet-processing pipeline to expose differentiated inference behaviors for different traffic or service states while preserving online responsiveness.

This challenge is not well addressed in existing literature. Prior work on anomaly detection, intrusion detection, and lightweight security inference mainly improves model accuracy, detection quality, or execution efficiency for specific packet-processing tasks~\cite{kitsune,survey-anomaly,survey-ids-2025,rplids-2024}. Research on SmartNICs, FPGAs, and programmable switches focuses on accelerating in-path inference or reducing the execution cost of network-resident processing~\cite{pigasus,taurus,azure-smartnic,bos,fenix}. Runtime-programmable data-plane systems further improve deployment flexibility and functional expressiveness~\cite{runtime-programmable-switches,ipsa}. However, these works do not explicitly study lightweight online model-switching within a shared forwarding framework for in-network artificial computing in emergency communications networks.

The key issue addressed in this paper is not only whether a single model can run efficiently on the packet path, but whether multiple lightweight inference behaviors can be supported and switched online at low cost. Binary Neural Networks (BNNs) are adopted because their compact parameter state and efficient SIMD mapping make them suitable for lightweight in-network execution and low-overhead model residency on commodity CPUs~\cite{binaryconnect,xnornet}.

In light of these observations, the primary contributions of this paper are as follows:
\begin{itemize}
    \item We formulate enhanced light model-switching for in-network artificial computing as a systems problem in emergency communications networks.
    \item We propose a shared execution framework that combines lightweight resident models, packet-granular metadata-driven model selection, and a unified forwarding path to realize responsive online semantic adaptation with low switching overhead.
    \item We implement the proposed design on a non-intrusive \mbox{eBPF/XDP + AF\_XDP} stack and show that it achieves lightweight online inference, preserves low switching overhead under resident-bank scaling, and supports packet-boundary model switching without wrong-verdict packets in the evaluated runs.
\end{itemize}

The remainder of this paper is organized as follows. Section~\ref{sec:design} details the proposed system design. Section~\ref{sec:eval} reports the comprehensive evaluation results. Finally, Sections~\ref{sec:conclusion} provide the  conclusion, respectively.

\section{System Design}
\label{sec:design}

\sys supports enhanced light model-switching for in-network artificial computing, allowing multiple model behaviors within the same forwarding path. The active model is selected online through packet metadata, enabling differentiated inference without modifying the path.

This design is built on three elements, namely a fixed packet representation, a resident model bank, and a shared inline executor. Throughout packet processing, the parser, the executor, and the forwarding logic remain unchanged, while only the selected model is updated through metadata resolution. The rest of this section describes these components in turn and then presents the prototype implementation on the \mbox{eBPF/XDP + AF\_XDP} stack.

\subsection{Architecture Overview}

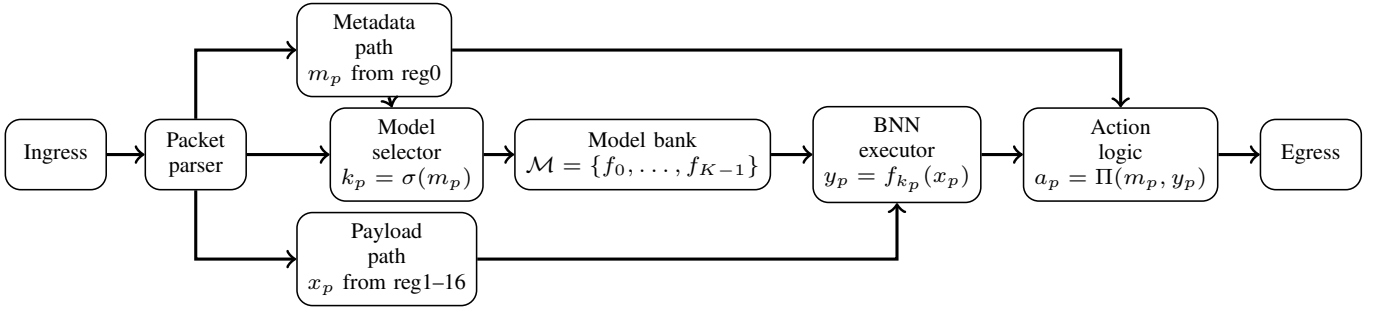
\begin{figure*}[htbp]
\centering
\resizebox{\linewidth}{!}{%
\begin{tikzpicture}[font=\scriptsize, node distance=0.34cm, scale=0.82, transform shape]
\tikzstyle{blk}=[draw, rounded corners, align=center, minimum height=0.78cm, minimum width=1.12cm]
\tikzstyle{smallblk}=[draw, rounded corners, align=center, minimum height=0.62cm, minimum width=1.20cm]
\node[blk] (ing) {Ingress};
\node[blk, right=0.42cm of ing] (parser) {Packet\\parser};
\node[smallblk, above right=0.24cm and 0.55cm of parser] (meta) {Metadata\\path\\$m_p$ from reg0};
\node[smallblk, below right=0.24cm and 0.55cm of parser] (payload) {Payload\\path\\$x_p$ from reg1--16};
\node[blk, right=0.92cm of parser] (selector) {Model\\selector\\$k_p=\sigma(m_p)$};
\node[blk, right=0.34cm of selector] (bank) {Model bank\\$\mathcal{M}=\{f_0,\dots,f_{K-1}\}$};
\node[blk, right=0.46cm of bank] (exec) {BNN\\executor\\$y_p=f_{k_p}(x_p)$};
\node[blk, right=0.46cm of exec] (action) {Action\\logic\\$a_p=\Pi(m_p,y_p)$};
\node[blk, right=0.46cm of action] (eg) {Egress};
\draw[->, thick] (ing) -- (parser);
\draw[->, thick] (parser) -- (selector);
\draw[->, thick] (parser) |- (meta);
\draw[->, thick] (parser) |- (payload);
\draw[->, thick] (meta) -- (selector);
\draw[->, thick] (selector) -- (bank);
\draw[->, thick] (bank) -- (exec);
\draw[->, thick] (payload.east) -| (exec.south);
\draw[->, thick] (meta.east) -| (action.north);
\draw[->, thick] (exec) -- (action);
\draw[->, thick] (action) -- (eg);
\end{tikzpicture}
}
\caption{Embedded BNN switch architecture. \texttt{reg0} carries metadata for model selection, \texttt{reg1}--\texttt{reg16} carry the payload blocks for inference, and the model bank enables lightweight online model switching across preloaded weight sets.}
\label{fig:system}
\end{figure*}

Fig.~\ref{fig:system} illustrates the packet path architecture of \sys. After ingress, each packet is separated into a metadata path and a payload path. Control information is extracted from \texttt{reg0} to drive model slot selection, while the payload region from \texttt{reg1} to \texttt{reg16} is assembled as the fixed input of the shared executor. The selected resident model is then applied to the payload, and the final forwarding action is determined jointly by the metadata and the inference result.

A defining property of this design is that light model-switching is achieved without modifying the forwarding pipeline itself. Across packets, the parser, the executor, and the forwarding logic remain unchanged, while only the referenced model slot is resolved differently through packet metadata. As a result, differentiated inference behaviors are realized as an online packet-path operation within a shared execution framework, rather than as a conventional reconfiguration event imposed on the path.

\subsection{Packet Formulation and Embedded BNN Execution}

To support lightweight online model switching at packet boundaries, a fixed packet representation is adopted in \sys. Each packet is represented as a 1088 byte sample composed of seventeen 64 byte register blocks. The first block, \texttt{reg0}, carries control metadata and configuration fields. The organization of these metadata fields is summarized in Table~\ref{tab:reg0}. In particular, the 4 byte Model Slot ID is extracted to determine the active model index $k_p$ for each packet. Through this arrangement, model selection is kept strictly outside the payload region consumed by the BNN executor.

\begin{table}[h]
\caption{\texttt{reg0} Metadata Organization in the Prototype}
\label{tab:reg0}
\centering
\begin{tabular}{lll}
\toprule
Field & Size & Function \\
\midrule
Model slot ID & 4 B & Selects $k_p$  \\
Format / version & 4 B & Guards parser compatibility  \\
Control / reserved & 8 B & Future packet actions  \\
Padding / spare metadata & 48 B & Kept outside BNN input  \\
\bottomrule
\end{tabular}
\end{table}

\begin{figure}[t]
\centering
\begin{tikzpicture}[node distance=0cm, font=\scriptsize]
\draw[fill=gray!20] (0,0) rectangle (1.1,0.5) node[pos=.5] {reg0};
\draw[fill=blue!10] (1.1,0) rectangle (2.2,0.5) node[pos=.5] {reg1};
\draw[fill=blue!10] (2.2,0) rectangle (3.3,0.5) node[pos=.5] {reg2};
\draw[fill=white] (3.3,0) rectangle (4.5,0.5) node[pos=.5] {$\dots$};
\draw[fill=blue!10] (4.5,0) rectangle (5.6,0.5) node[pos=.5] {reg15};
\draw[fill=blue!10] (5.6,0) rectangle (6.7,0.5) node[pos=.5] {reg16};

\node[above] at (0.55, 0.5) {Meta};
\node[above] at (3.9, 0.5) {BNN Payload (1024B)};

\draw[->, dashed, gray] (0.55, 0) -- (0.55, -0.7) node[below, font=\tiny, yshift=-2pt] {Slot Selection};
\draw[->, thick, blue!50] (1.65, 0) -- (1.65, -0.7);
\draw[->, thick, blue!50] (2.75, 0) -- (2.75, -0.7);
\draw[->, thick, blue!50] (5.05, 0) -- (5.05, -0.7);
\draw[->, thick, blue!50] (6.15, 0) -- (6.15, -0.7);

\node at (3.9, -0.35) {Direct Register Mapping};

\draw[thick, fill=orange!20] (1.1,-1.2) rectangle (2.2,-0.7) node[pos=.5] {ZMM0};
\draw[thick, fill=orange!20] (2.2,-1.2) rectangle (3.3,-0.7) node[pos=.5] {ZMM1};
\draw[thick, fill=white] (3.3,-1.2) rectangle (4.5,-0.7) node[pos=.5] {$\dots$};
\draw[thick, fill=orange!20] (4.5,-1.2) rectangle (5.6,-0.7) node[pos=.5] {ZMM14};
\draw[thick, fill=orange!20] (5.6,-1.2) rectangle (6.7,-0.7) node[pos=.5] {ZMM15};

\node[below] at (3.9, -1.2) {x86 AVX-512 Registers (512-bit each)};
\end{tikzpicture}
\caption{The hardware aligned mapping between packet registers and AVX 512 ZMM units. By partitioning the payload into 64 byte blocks, \sys achieves direct vector loading for sub microsecond inference.}
\label{fig:mapping}
\end{figure}
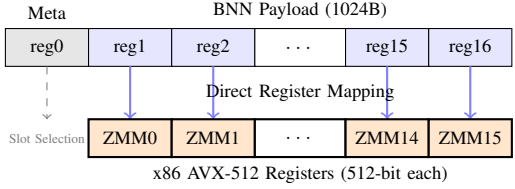

The remaining sixteen blocks, from \texttt{reg1} to \texttt{reg16}, carry the 1024 byte payload presented to the inline executor. This layout is chosen in accordance with the x86 AVX 512 execution width. As shown in Fig.~\ref{fig:mapping}, each 64 byte block matches the 512 bit width of one ZMM register. As a result, the payload can be loaded through a regular register aligned mapping, and additional memory alignment penalties are avoided. The packet is therefore written in the form $p=(m_p, x_p)$, where $m_p$ denotes the metadata and $x_p$ denotes the payload.

Binary inference is performed by the executor according to
\begin{equation}
h_p = \mathrm{sign}(W_k^{(1)}x_p + b_k^{(1)}), \quad y_p = W_k^{(2)}h_p + b_k^{(2)} .
\end{equation}
Because binary activations are used, the execution path can be expressed through bitwise operations together with popcount style accumulation. A regular memory access pattern is thereby preserved, and the compute cost per packet remains bounded and predictable.

\subsection{Weight Memory Organization and Lightweight Online Model Switching}

Multiple BNN weight sets are preloaded into a resident model bank $\mathcal{M}$, where each slot $f_k$ stores a complete parameter set:
\begin{equation}
\mathcal{M}=\{f_0,f_1,\dots,f_{K-1}\},
\end{equation}
\begin{equation}
f_k=\left(W_k^{(1)},b_k^{(1)},W_k^{(2)},b_k^{(2)}\right).
\end{equation}

In the current prototype, each slot stores the complete parameter set together with the executor-side metadata required at runtime. All resident models share the same input representation and the same execution interface in the present design, while their weights and biases differ across slots. All slots are loaded during initialization and remain resident at fixed memory locations throughout runtime. Consequently, model switching is reduced to slot selection over already resident model objects, rather than being implemented through online weight delivery or forwarding-path reconfiguration.

For an incoming packet $p=(m_p,x_p)$, the model selection field is extracted from $m_p$ to compute the slot index $k_p$, the inference result $y_p$, and the final forwarding action $a_p$ as follows:
\begin{equation}
k_p = \sigma(m_p),
\end{equation}
\begin{equation}
y_p = f_{k_p}(x_p),
\end{equation}
\begin{equation}
a_p = \Pi(m_p, y_p).
\end{equation}

In this process, the selected resident model is applied to the payload, and the final forwarding action is derived accordingly. This organization captures the central design property of \sys. Model switching is realized through slot selection rather than forwarding-path mutation. Across packets, the parser, the executor, and the forwarding logic remain unchanged, while only the referenced resident model differs according to the resolved slot index. In this sense, lightweight online model switching is realized as an in-path runtime operation within the shared execution framework, without being elevated into a reconfiguration event.

Because all model slots share the same input format and the same execution interface, runtime switching does not alter the parser, the executor pipeline, or the forwarding logic. Only the slot index $k_p$ is updated for each packet. As a result, lightweight online model switching can be performed at packet granularity while traffic continues to traverse the same forwarding path.

\begin{figure}[t]
\centering
\begin{tikzpicture}[font=\scriptsize, node distance=0.25cm]
\tikzstyle{blk}=[draw, rounded corners, align=center, minimum height=0.62cm, minimum width=1.12cm]
\tikzstyle{slot}=[draw, align=center, minimum height=0.56cm, minimum width=1.50cm]
\node[blk] (meta) {\texttt{reg0}\\slot field};
\node[blk, right=0.45cm of meta] (selector) {slot selector\\$k_p$};
\node[slot, right=0.7cm of selector, yshift=0.75cm] (s0) {slot 0\\resident weights};
\node[slot, right=0.7cm of selector] (s1) {slot 1\\resident weights};
\node[slot, right=0.7cm of selector, yshift=-0.75cm] (sk) {slot $K\!-\!1$\\resident weights};
\node[blk, right=0.7cm of s1] (exec) {shared BNN\\executor};
\draw[->, thick] (meta) -- (selector);
\draw[->, thick] (selector) -- (s0);
\draw[->, thick] (selector) -- (s1);
\draw[->, thick] (selector) -- (sk);
\draw[->, thick] (s0.east) -- (exec.west);
\draw[->, thick] (s1.east) -- (exec.west);
\draw[->, thick] (sk.east) -- (exec.west);
\end{tikzpicture}
\caption{Resident model bank layout. Multiple weight sets are kept resident in memory behind a shared executor, and lightweight online model switching is realized by resolving the referenced slot.}
\label{fig:modelbank}
\end{figure}
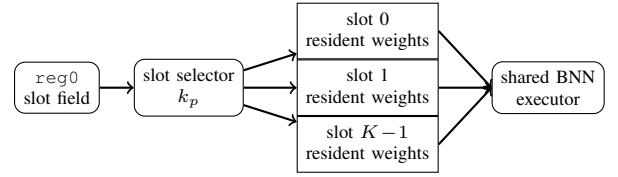

Fig.~\ref{fig:modelbank} illustrates the resident model bank used in the prototype. Each slot corresponds to a fixed in memory parameter object with the same structural layout. As a result, data path switching is reduced to resolving the referenced resident slot, rather than rebuilding model state or reconfiguring the forwarding path.

\begin{algorithm}[h]
\caption{Packet-Path Processing with Lightweight Online Model Selection}
\label{alg:hotswitch}
\begin{algorithmic}[1]
\Require packet $p=(m_p,x_p)$, model bank $\mathcal{M}$
\State parse slot metadata from \texttt{reg0}
\State compute slot index $k_p \gets \sigma(m_p)$
\State  resolve resident slot $k_p$ and fetch model $f_{k_p}$ from $M$
\State compute score $y_p \gets f_{k_p}(x_p)$
\State derive action $a_p \gets \Pi(m_p,y_p)$
\State emit packet according to $a_p$
\end{algorithmic}
\end{algorithm}

\subsection{Prototype Realization and Memory Footprint}

The proposed architecture is realized on a Linux native packet path built from eBPF/XDP and AF\_XDP. At ingress, packets are steered by XDP into a user space forwarding path backed by AF\_XDP. Within this path, both the resident model bank and the shared AVX 512 BNN executor are hosted by the same forwarder. For each packet, the slot field is read from \texttt{reg0}, the corresponding resident slot is resolved, and the same executor is applied to the fixed 1024 byte payload representation. In this way, the single pipeline property of \sys\ is preserved. Switching neither creates a new execution path nor instantiates another executor. Only the referenced resident slot is changed at runtime.

Under this implementation, lightweight online model switching remains strictly below the level of forwarding-path reconfiguration. Traffic is neither redirected to another service nor coupled to online weight loading. In the current prototype, the post inference action stage is intentionally kept simple, so that the evaluation can isolate whether different resident models produce distinct and observable packet behaviors on the same forwarding path.

The resident footprint of the prototype remains modest on commodity CPUs. In the current implementation, each \texttt{h32} weight file used in the experiments occupies 32932 bytes on disk. Accordingly, the two slot online prototype keeps about 64.3~KB of resident weights, whereas the 16 slot scaling microbenchmark keeps about 514.6~KB. The corresponding footprint is summarized in Table~\ref{tab:slots}.

\begin{table}[t]
\caption{Resident Weight Footprint in the Prototype}
\label{tab:slots}
\centering
\begin{tabular}{lcc}
\toprule
Resident bank & Slots & Weight footprint \\
\midrule
Online continuity prototype & 2 & 65864 B \\
Scaling microbenchmark & 16 & 526912 B \\
\bottomrule
\end{tabular}
\end{table}

The execution cost of the hot switching path can also be described structurally. Let $d$ denote the input dimension in bits, and let $h$ denote the hidden layer width. Under the fixed format packet path used in this prototype, one packet requires one slot lookup, one binary hidden layer pass over $d \times h$, and one output layer pass over $h$. Parsing and slot selection therefore remain constant cost operations in the evaluated design, while the model dependent computation scales only with the selected resident BNN. This characterization keeps the per packet switching cost regular and predictable in the evaluated regime. At the same time, it should be noted that the present analysis does not yet characterize larger resident banks that may impose stronger pressure on the cache hierarchy.

\section{Evaluation}
\label{sec:eval}

\sys is evaluated here as an in-network artificial computing framework with enhanced light model-switching, rather than as a detector design study. Accordingly, this section focuses on three questions. First, whether inline BNN execution is sufficiently lightweight for online deployment. Second, whether metadata-driven model selection can induce distinct packet-processing behaviors while the forwarding path remains shared. Third, whether lightweight online model switching preserves low overhead and forwarding continuity as the resident model bank is extended beyond the two-slot setting. Unless otherwise stated, the fixed single-model path is used as the baseline operating mode.

\subsection{Experimental Setup}

\sys is instantiated with malicious traffic recognition as a concrete packet processing workload, while the evaluation remains focused on switching behavior rather than detector quality. Packet samples are derived from IoT-23~\cite{iot23} and are mapped into the fixed 1088 byte representation used by \sys. The same \texttt{h32} BNN structure is used throughout all experiments. Different model slots share the same input format, the same executor, and the same forwarding path, while differing only in their resident weight sets.

\begin{table}[t]
\caption{Experimental Setup Summary}
\label{tab:exp_setup}
\centering
\begin{tabular}{p{2.7cm}p{5.2cm}}
\toprule
Item & Configuration \\
\midrule
Workload & Malicious traffic recognition \\
Dataset & IoT-23~\cite{iot23} \\
Packet format & Fixed 1088 byte representation, with \texttt{reg0} for metadata and \texttt{reg1}--\texttt{reg16} for 1024 byte payload \\
Model structure & Shared \texttt{h32} BNN structure in all experiments \\
Two slot setup & One shared executor and two resident slots with different weight sets only \\
Training split & Training groups: \texttt{20-1}, \texttt{21-1}, \texttt{33-1}, \texttt{36-1}, \texttt{43-1}, \texttt{48-1}; validation groups: \texttt{35-1}, \texttt{42-1} \\
Slot 0 & Recall oriented model, trained with \texttt{pos\_weight=4.0}, selected by recall \\
Slot 1 & Precision oriented model, trained with \texttt{pos\_weight=0.5}, selected by precision \\
Slot loading & Both weight sets are preloaded before runtime \\
16 slot setup & The same two weight sets are alternated across 16 resident slots \\
Scaling traces & Fixed, round robin, random, and hotspot slot access traces \\
\bottomrule
\end{tabular}
\end{table}

Under this setup, the two-slot experiment is used to test whether resident model selection can produce distinguishable packet-processing behaviors and lightweight boundary switching on a shared forwarding path. The 16-slot experiment is used only to evaluate whether model-selection cost remains stable as resident-bank cardinality increases, rather than to represent 16 distinct application behaviors. Online switching is evaluated with a deterministic 64 packet stream for boundary correctness and with a longer 8192 packet stream under the same slot transition at larger scale.

\subsection{Runtime Cost}

The runtime overhead of \sys is first evaluated to determine whether lightweight online model switching can remain on the packet path without violating the forwarding budget. As shown in Fig.~\ref{fig:runtime}, the \texttt{h32} executor completes one inference in 0.528~$\mu$s and sustains 1.894~Mpps on one pinned core. For reference, this corresponds to 15.52~Gbps under a 1024 byte payload interpretation and 22.73~Gbps under a 1500 byte packet length interpretation.

Fig.~\ref{fig:runtime} further breaks the runtime cost into slot selection, inference, and end to end packet path latency. Pure slot selection costs 0.005~$\mu$s, whereas pure BNN inference costs 0.528~$\mu$s. The total packet path latency, including AF\_XDP I/O, packet parsing, slot selection, inference, and forwarding logic, is 0.894~$\mu$s. These results show that the runtime overhead introduced by resident slot selection is negligible relative to inference and remains compatible with deployment on commodity CPUs.

A dedicated microbenchmark is then used to examine whether this property is preserved as resident bank cardinality increases from 2 to 16 slots. The same two \texttt{h32} weight sets are alternated across resident slots so that slot selection cost can be isolated from parser and executor changes. Correct slot selection is preserved for all 16 slot IDs. As shown in Fig.~\ref{fig:scaling}, across fixed, round robin, random, and hotspot access traces, the measured slot selection cost remains tightly clustered around 0.0037~$\mu$s for both 2 slot and 16 slot banks. The combined selection plus inference latency remains within 0.67~$\mu$s to 0.92~$\mu$s across the same regimes, where the larger variation is attributable to access pattern dependent runtime behavior rather than to slot selection itself. These results indicate that resident bank cardinality can be increased to 16 without materially changing the cost of the slot selection path.

\begin{figure}[t]
\centering
\begin{tikzpicture}
\begin{axis}[
    ieeeplot,
    height=3.8cm,
    ybar,
    bar width=12pt,
    ymin=0,
    ymax=1.0,
    ylabel={Latency ($\mu$s)},
    symbolic x coords={Select,Infer,End-to-end},
    xtick=data,
    enlarge x limits=0.20,
    xticklabel style={font=\scriptsize, rotate=15, anchor=east}
]
\addplot+[draw=black, fill=blue!45] coordinates {(Select,0.005) (Infer,0.528) (End-to-end,0.894)};
\end{axis}
\end{tikzpicture}
\caption{Runtime breakdown of slot selection, inline \texttt{h32} inference, and end to end packet path latency.}
\label{fig:runtime}
\end{figure}
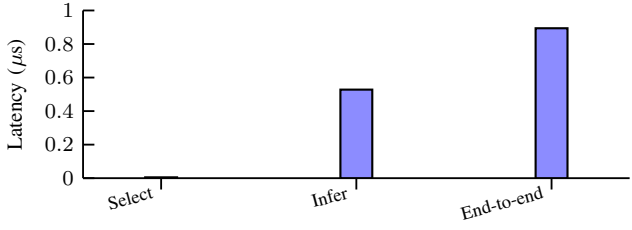

\begin{figure*}[t]
\centering
\begin{minipage}[t]{0.48\textwidth}
\centering
\begin{tikzpicture}
\begin{axis}[
    ieeeplot,
    width=\linewidth,
    height=4.0cm,
    ybar,
    bar width=10pt,
    ymin=0,
    ymax=0.005,
    ylabel={Slot selection latency ($\mu$s)},
    symbolic x coords={Fixed,RR,Rand,Hot},
    xtick=data,
    xticklabel style={font=\scriptsize, rotate=20, anchor=east},
    legend style={at={(0.5,-0.18)}, anchor=north, legend columns=2},
    enlarge x limits=0.16
]
\addplot+[draw=black, fill=blue!40] coordinates {(Fixed,0.00375068) (RR,0.00373028) (Rand,0.00372933) (Hot,0.00374444)};
\addplot+[draw=black, fill=orange!65] coordinates {(Fixed,0.00371203) (RR,0.00372434) (Rand,0.00370111) (Hot,0.00371265)};
\legend{2-slot,16-slot}
\end{axis}
\end{tikzpicture}
\end{minipage}\hfill
\begin{minipage}[t]{0.48\textwidth}
\centering
\begin{tikzpicture}
\begin{axis}[
    ieeeplot,
    width=\linewidth,
    height=4.0cm,
    ybar,
    bar width=10pt,
    ymin=0,
    ymax=1.0,
    ylabel={Select + Inference Latency($\mu$s)},
    symbolic x coords={Fixed,RR,Rand,Hot},
    xtick=data,
    xticklabel style={font=\scriptsize, rotate=20, anchor=east},
    enlarge x limits=0.16
]
\addplot+[draw=black, fill=blue!40] coordinates {(Fixed,0.687227) (RR,0.895463) (Rand,0.779513) (Hot,0.779394)};
\addplot+[draw=black, fill=orange!65] coordinates {(Fixed,0.669576) (RR,0.914953) (Rand,0.896965) (Hot,0.832076)};
\end{axis}
\end{tikzpicture}
\end{minipage}
\caption{Resident-bank scaling from 2 to 16 slots under different slot-access patterns. Left: Slot selection latency remains nearly unchanged. Right: Select plus inference latency is dominated by access-pattern-dependent runtime behavior.}
\label{fig:scaling}
\end{figure*}

\subsection{Slot-Conditioned Behavior}

\begin{figure}[t]
\centering
\begin{tikzpicture}
\begin{axis}[
    ieeeplot,
    height=3.9cm,
    ybar,
    bar width=10pt,
    ymin=0,
    ymax=1.05,
    ylabel={Metric Value},
    symbolic x coords={Precision,Recall,F1},
    xtick=data,
    legend style={at={(0.5,-0.18)}, anchor=north, legend columns=2},
    enlarge x limits=0.22,
    legend cell align=left
]
\addplot+[draw=black, fill=blue!40] coordinates {(Precision,0.6244) (Recall,1.0) (F1,0.7688)};
\addplot+[draw=black, fill=orange!65] coordinates {(Precision,0.8298) (Recall,0.0095) (F1,0.0188)};
\legend{Slot 0 (recall), Slot 1 (precision)}
\end{axis}
\end{tikzpicture}
\caption{Slot conditioned behavior on the same forwarding path, shown in terms of precision, recall, and F1 score.}
\label{fig:slotbehavior}
\end{figure}
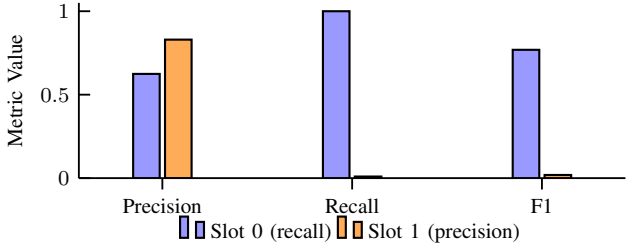

Metadata driven slot selection is next examined for its ability to induce distinct packet processing behaviors on the same forwarding path. The two resident slots are intentionally configured to exhibit different decision tendencies under the same workload. Slot~0 is recall oriented, whereas Slot~1 is precision oriented. As shown in Fig.~\ref{fig:slotbehavior}, this difference is clearly reflected in the resulting precision, recall, and F1 scores. Slot~0 produces high recall, while Slot~1 produces substantially higher precision. These results are sufficient to show that different resident slots can expose distinguishable packet handling behaviors on the same shared forwarding path.

This effect is also directly observable at the single sample level. With the payload held fixed, changing only the slot selection field in \texttt{reg0} changes the output score from 1.98715 under Slot~0 to -0.0181384 under Slot~1. This result confirms that packet behavior can be altered solely through slot choice, while the forwarding path itself remains unchanged.

\subsection{Switching Continuity}

Lightweight online model switching is first tested in terms of forwarding continuity. A deterministic 64-packet trace is used to examine whether model switching can be completed correctly at a packet boundary on the shared forwarding path. The first half of the stream carries \texttt{reg0=0} and consistently selects Slot~0, while the second half carries \texttt{reg0=1} and consistently selects Slot~1. The transition occurs exactly at the packet boundary, from source port 47031 to 47032.

An 8192 packet run is then used to examine continuity at larger scale. This run uses a user space replay harness that spaces packet emissions by 10~$\mu$s. The pacing is intentional, so that per packet switching continuity can be examined without being dominated by PCIe or driver side batching artifacts. In this run, the forwarder processes 8131 of 8192 packets online, or 99.26\%. Trace inspection attributes the missing 61 packets to the initial warm up prefix of the replay rather than to the switching boundary. Across all observed packets, zero wrong slot hits and zero wrong forwarding verdicts are recorded, and all 4096 Slot~1 packets in the sink phase are delivered successfully.

From the timing perspective, Table~\ref{tab:continuity} summarizes the continuity statistics on the 8192-packet run. The measured boundary gap of 95.58~$\mu$s remains close to the overall median gap of 93.03~$\mu$s, while the forwarding rate within a 512-packet window is 10.49~kpps before the switch and 10.85~kpps after it. These results support the conclusion that, in the evaluated run, lightweight online model switching remains correct at the packet boundary and does not create a visible continuity break on the forwarding path.

\subsection{Comparison with Online Control-Plane Replacement}

To compare lightweight online model switching with online control-plane replacement, a real control-plane update experiment is executed on the same 8192-packet workload. In that run, the forwarder starts with only Slot~0 active, and the control plane sends the Slot~1 weight file over a Unix control socket only after the packet stream crosses the switching boundary. Under the same operation-level notion of switching cost used in Fig.~\ref{fig:runtime}, the control-plane switch latency from update send start to effective is 484.90~$\mu$s. Because triggering starts only after boundary detection, the boundary-to-effective window expands to 8479.45~$\mu$s, during which 99 post-boundary packets are still processed under Slot~0 and therefore produce 99 wrong-model and 99 wrong-verdict events.

This contrast highlights the role of resident preloading in \sys. Under a unified switching definition, lightweight resident switching costs 0.005~$\mu$s, while online control-plane switching costs 484.90~$\mu$s. The control update result should therefore not be interpreted as a second implementation of the same mechanism. Instead, it serves as a comparison point showing both the higher switching cost and the resulting post-boundary error window when the new model is not already resident on the forwarding path.

Table~\ref{tab:controlcompare} summarizes the direct comparison between lightweight resident switching and online control-plane replacement in terms of switch latency and wrong-packet count.

\begin{table}[t]
\caption{Switching continuity statistics on the 8192-packet run}
\label{tab:continuity}
\centering
\begin{tabular}{lcc}
\toprule
Metric & Before / Median & After / Boundary \\
\midrule
Packet gap ($\mu$s) & 93.03 & 95.58 \\
Forwarding rate (kpps) & 10.49 & 10.85 \\
\bottomrule
\end{tabular}
\end{table}

\begin{table}[t]
\caption{Comparison between lightweight resident switching and online control-plane replacement}
\label{tab:controlcompare}
\centering
\begin{tabular}{lcc}
\toprule
Method & Switch latency ($\mu$s) & Wrong packets \\
\midrule
Lightweight resident switching & 0.005 & 0 \\
Online control-plane replacement & 484.896 & 99 \\
\bottomrule
\end{tabular}
\end{table}

\section{Conclusion}
\label{sec:conclusion}

We present an in-network artificial computing framework with lightweight model-switching for emergency communications networks, using packet metadata to select models without modifying the forwarding path. This enables lightweight online switching as an in-path operation, rather than a control-plane replacement.

The evaluation shows that this design is practical on commodity x86 hardware. The \texttt{h32} executor sustains 1.894~Mpps with 0.528~$\mu$s inference latency, while model switching adds only 0.005~$\mu$s of extra cost. Different resident models produce distinguishable packet-processing behaviors on the same forwarding path, and lightweight online switching is completed without visible continuity break in the evaluated runs.

The control-plane update experiment further shows why resident preloading is important for this design. Under the same switching metric, lightweight resident switching costs 0.005~$\mu$s, whereas online control-plane switching costs 484.90~$\mu$s and produces a non-zero post-boundary wrong-packet window. These results indicate that the proposed design provides a practical basis for lightweight in-network artificial computing and responsive online semantic adaptation in emergency communications networks.

\section*{Acknowledgment}

This work was supported by the National Key Research and Development Program of China (No. 2024YFE0200302).

\bibliographystyle{IEEEtran}
\bibliography{refs}

\end{document}